\documentclass[11pt]{article}
\usepackage{color}
\usepackage[body={16cm,23cm}]{geometry}
\usepackage{amssymb,amsmath,amsthm,amsopn}
\usepackage{amsfonts}
\usepackage[mathscr]{eucal}
\usepackage{enumerate}
\usepackage{graphicx}

\makeatletter
\@addtoreset{equation}{section}
\makeatother

\newcommand{\ds}{\displaystyle}

\newcommand{\eop}{\boldsymbol{e}}
\newcommand{\fop}{\boldsymbol{f}}
\newcommand{\hop}{\boldsymbol{h}}
\newcommand{\ii}{\mathsf{i}}

\renewcommand{\author}[1]{\large\rm #1\\ \bigskip}
\newcommand{\address}[1]{{\normalsize\it #1\\}\bigskip}
\renewcommand{\title}[1]{\bigskip\bigskip\Large\bf #1\bigskip\bigskip\\}

\begin{document}

\vglue 2cm

\begin{center}

\title{On eigenstates for some $sl_2$ related Hamiltonian}

\vspace{.5cm}

\author{Fahad M. Alamrani}

\vspace{.5cm}

\address{
Faculty of Education Science Technology \& Mathematics,
University of Canberra, Bruce ACT 2601, Australia.,\\
Department of Mathematics, Faculty of Science, 
University of Tabuk, P.O. BOX 741, Tabuk 7149.}

\abstract{
In this paper we consider the stationary Schr\"odinger equation for a self-conjugated Hamiltonian $\ds\boldsymbol{H}\; =\; \frac{\eop+\fop}{\ii}$, where $\eop$ and $\fop$ is an anti-unitary pair of the canonical Cartan "creating" and "annihilation" operators for the classical Lie algebra $sl_2$ taken in the representation with "the lowest weight equals to $1$''. In this paper we prove that this operator has the continuous spectrum. Construction of eigenstates for $\boldsymbol{H}$ is given in details. 
}

\end{center}


\vspace{1cm}

\section{Introduction}

This paper will deal with the representation theory of  the classical Lie algebra \cite{JH}. We will consider the Lie algebra $sl_2$ in a certain infinitely dimensional representation corresponding to the lowest weight $1$. The representation module is equivalent to the Fock Space representation of the quantum oscillator \cite{VJ}. The "creating" and "annihilation" operators $\eop$ and $\fop$ are anti-unitary, so that the operator $\boldsymbol{H}\; =\; \frac{1}{\ii} (\eop+\fop)$ is Hermitian, and therefore it can be interpreted as a Hamiltonian for a certain Quantum Mechanical system. This Hamiltonian is related to a Hamiltonian considered in \cite{SM,KS} in the limit $q=1$ (Note, the regime $q=1$ was not considered in \cite{SM,KS}).

This paper organised as follows. In section $2$ we fix the proper representation of $sl_2$ and rewrite the stationary Schr\"odinger equation as a linear recursion with non-constant coefficients. Section $3$ is devoted to the analysis of the recursion equations. Its asymptotic is discussed in section $4$. Section $5$ contains discussion and conclusion. 

\section{Formulation of the problem}

We consider the algebra $sl_2$ generated by three operators $\eop,\fop,\hop$ satisfying the three fundamental commutation relations \cite{JH}.
\begin{equation}
[\eop,\fop]\;=\;\hop\;,\quad [\hop,\eop]\;=\;2\eop\;,\quad [\hop,\fop]\;=\;-2\fop\;.
\end{equation}
Let $\mathfrak{F}$ stands for the Fock Space,
\begin{equation}
\mathfrak{F}\;=\;\mathrm{Span}\biggl\{\; |\, n\rangle\;,\quad n\in\mathbb{Z}_{n\geq 0}\;\biggr\}\;.
\end{equation}
The map
\begin{equation}
\eop\;\stackrel{\pi}{\mathop{\rightarrow}}\;\pi(\eop)\;\in\;\mathrm{End}\,(\mathfrak{F})\;,\quad \textrm{etc.,}
\end{equation}
we define as
\begin{equation}\label{therep}
\eop\;|n\rangle\;=\;|n+1\rangle\; \ii (n+1)\;,\quad \fop\;|n\rangle\;=\;|n-1\rangle\; \ii n\;,\quad
\hop\;|n\rangle \;=\; |n\rangle \; (2n+1)\;,\quad n\in\mathbb{Z}_{n\geq 0}\;,
\end{equation}
where for shortness we use notation $\eop$ instead of $\pi(\eop)$, etc.
Our representation (\ref{therep}) is the representation with the lowest weight $1$, 
\begin{equation}
\hop |0\rangle \;=\; |0\rangle\;.
\end{equation}  
(in Physics this is called "spin $=  -1/2$ representation''). The Fock co-module is defined by 
\begin{equation}\label{comodule}
\langle n|n'\rangle \;=\; \delta_{n,n'}\;,\quad n,n'\geq 0\;.
\end{equation}
An essential feature of our paper is that this representation not unitary:
\begin{equation}
\eop^\dagger \;=\; - \fop\;,
\end{equation}
where the ``dagger'' means the Hermitian conjugation.
Subject of our interest is self-conjugated unbounded Hamiltonian
\begin{equation}
\boldsymbol{H}\;=\;\frac{\eop+\fop}{\ii}\;,
\end{equation}
and the stationary Schr\"odinger equation for it,
\begin{equation}\label{Sch-e}
\boldsymbol{H}\;|\psi\rangle \;=\; |\psi\rangle \;E\;.
\end{equation}
In what follows, we will study the structure of $|\psi\rangle$ for any $E\;\in\;\mathbb{R}$ and deduce that our Hamiltonian has continuous spectrum.

\section{Analysis of the recursion}

We will use the Dirac  notations for $\langle \textrm{bra}|$ and $|\textrm{ket}\rangle$ vectors.
In components,
\begin{equation}
\psi_n\;=\;\langle n|\psi\rangle\;,
\end{equation}
where $\langle n|$ is a state of Fock co-module, cf. (\ref{comodule}), and $|\psi\rangle$ is a required wavefunction.
The stationary Schr\"odinger equation (\ref{Sch-e}) in components reads
\begin{equation}\label{recursion}
(n+1)\;\psi_{n+1} \;+\; n\; \psi_{n-1} \;=\; E\;\psi_n\;,
\end{equation}
where we assume 
\begin{equation}
\psi_0 \;=\; 1\qquad \forall\;\; E\;\in\;\mathbb{R}\;.
\end{equation}
Our aim now is to understand the asymptotic behaviour of $\psi_n$ when $n\to\infty$. Since $E$ for now is only one free parameter, we assume implicitly
\begin{equation}
|\psi\rangle\;=\;|\psi_E\rangle\;,\quad \psi_n\;=\;\psi_n(E)\;.
\end{equation}

Recursion (\ref{recursion}) can be identically rewritten in matrix form \cite{SM,KS}:
\begin{equation}
(\psi_n,\psi_{n+1}) \;=\; (\psi_{n-1},\psi_n)\;\cdot\; L_{n+1}\;,
\end{equation}
where
\begin{equation}
L_n\;=\;\left(\begin{array}{cc}
\ds 0 & \ds -1+\frac{1}{n} \\
\ds 1 & \ds \frac{E}{n}\end{array}\right)\;.
\end{equation}
Thus,
\begin{equation}\label{product}
(\psi_{n-1},\psi_n) \;=\; (0,1)\;L_1\cdot L_2 \cdots L_{n-1}\cdot L_n\;.
\end{equation}
Since 
\begin{equation}
L_\infty \;=\; \left(\begin{array}{cc} 0 & -1 \\ 1 & 0 \end{array}\right)\;,\quad L_\infty^4\;=\;1\;,
\end{equation}
we expect $mod\;\;4$ pattern for $\psi_n$. Diagonalising matrix $L_n$\;,
\begin{equation}
L_n\;=\;P_n^{-1} \; \left(\begin{array}{cc} \lambda_n & 0 \\ 0 & \overline{\lambda_n} \end{array}\right)\; P_n^{}\;,
\end{equation}
where 
\begin{equation}
\lambda_n\;=\;\ii \left(\sqrt{1-\frac{1}{n} - \frac{E^2}{4n^2}}-\ii\frac{E}{2n}\right)\;=\;
\ii \sqrt{1-\frac{1}{n}}\; \exp\left\{ -\ii \arcsin\frac{E}{2\sqrt{n(n-1)}}\right\}\;,
\end{equation}
and
\begin{equation}
P_n^{}P_{n+1}^{-1}\;=\;1+\frac{1}{2n^2}\left(\begin{array}{cc} 0 & 0 \\ -E & 1\end{array}\right) + \mathcal{O}(1/n^3)\;,
\end{equation}
one can deduce the following asymptotic straightforwardly from (\ref{product}):
\begin{equation}\label{asymptotic}
\psi_n(E)\;=\;\frac{A_n(E)}{\sqrt{n}}\; \cos\left( \frac{E}{2}\log n - \frac{\pi n}{2} + \varphi_n(E)\right)\;,\quad n \gg1\;.
\end{equation}
Intensive numerical computations allow one to conclude that the sequences $A_n(E)$ and $\varphi_n(E)$ smoothly converge to $A(E)$ and $\varphi(E)$ when $n\to\infty$. Therefore, we can postulate the $1/n$ expansion for $A_n$ and $\varphi_n$:
\begin{equation}\label{sequences}
A_n(E)\;=\; A(E)\;(1+\frac{\delta_1}{n}+\frac{\delta_2}{n^2}+\cdots)\;,\quad \varphi_n(E)\;=\;\varphi(E) + \frac{\epsilon_1}{n}+\frac{\epsilon_2}{n^2}+\cdots
\end{equation}
with some $n$-independent coefficients 
\begin{equation}
\delta_j\;=\;\delta_j(E)\;,\quad \epsilon_j=\epsilon_j(E)\;,\quad j\;\geq\;1\;.
\end{equation}
Values of $\delta_j, \;\epsilon_j$ must follow from (\ref{recursion}). In what follows, let us combine all correction terms in (\ref{sequences}) into
\begin{equation}\label{corr}
\delta(n,E)\;=\;\sum_{j=1}^\infty \frac{\delta_j(E)}{n^j}\;,\qquad
\epsilon(n,E)\;=\;\sum_{j=1}^\infty \frac{\epsilon_j(E)}{n^j}\;.
\end{equation}
To get these values, let us substitute (\ref{asymptotic}) into (\ref{recursion}). To do this in convenient way, let us introduce
\begin{equation}\label{Phi}
\Phi_n\;=\;\frac{E}{n}\log_n -\frac{\pi n}{2}+\varphi_n ;,\quad
\Phi_{n+1}\;=\;\Phi_n-\frac{\pi}{2}+\alpha_n;,\quad
\Phi_{n-1}\;=\;\Phi_n+\frac{\pi}{2}-\alpha'_n\;.
\end{equation}
The values of $\alpha_n$ and $\alpha'_n$ are then given by
\begin{eqnarray}\label{alpha}
\alpha_n\;=\;\Phi_{n+1}-\Phi_n+\frac{\pi}{2}\;=\;\frac{E}{2}\log_{(n+1)}+\varphi_{n+1}-\frac{E}{2}\log_n-\varphi_n\nonumber\\
\;=\;\frac{E}{2}\log(1+\frac{1}{n}) 
+\epsilon_1 (\frac{1}{n+1}-\frac{1}{n})+\epsilon_2 (\frac{1}{(n+1)^2}-\frac{1}{n^2})+\cdots 
\end{eqnarray}
and similarly for $\alpha_n'$. Let further
\begin{equation}
\frac{1}{n}=x \quad \Rightarrow \quad \frac{1}{n+1}=\frac{x}{1+x}=\sum_{j=1}^\infty (-)^{j+1} x^j \quad \textrm{etc.,}
\end{equation}
so that $1/n$-expansion becomes $x$-expansion. Then,
\begin{eqnarray}\label{alphaexp}
\alpha_n\;=\;\frac{E}{2}\log(1+x)+\epsilon_1(\frac{x}{1+x}-x)+\epsilon_2(\frac{x^2}{(1+x)^2}-x^2)+\cdots \nonumber\\
\;=\;\frac{E}{2}x-(\frac{E}{4}+\epsilon_1)x^2+(\frac{E}{6}+\epsilon_1-2\epsilon_2)x^3+\mathcal{O}(x^{4}).
\end{eqnarray}
Value of $\alpha'_n$ have similar structure. 

Now we can use (\ref{Phi},\ref{alpha} and \ref{alphaexp}) in (\ref{asymptotic} and \ref{recursion}):
\begin{equation}
\begin{array}{l}
\ds
\psi_n\;=\;\frac{A_n}{\sqrt{n}}\cos(\Phi_n)\;,\\
\ds \psi_{n+1}\;=\;\frac{A_{n+1}}{\sqrt{n+1}}\cos(\Phi_n-\frac{\pi}{2}+\alpha_n)\;=\;\frac{A_{n+1}}{\sqrt{n+1}}(\sin \Phi_n \cos \alpha_n+\cos \Phi_n \sin \alpha_n)\\
\ds \psi_{n-1}\;=\;\frac{A_{n-1}}{\sqrt{n-1}}(-\sin \Phi_n \cos \alpha'_n+\cos \Phi_n \sin \alpha'_n)\;.
\end{array}
\end{equation}
Equation (\ref{recursion}) can be written as 
\begin{equation}\label{coefficients}
\begin{array}{l}
\ds \cos \Phi_n \left[ (n+1)\frac{A_{n+1}}{\sqrt{n+1}}\sin \alpha_n+ n\frac{A_{n-1}}{\sqrt{n-1}}\sin \alpha'_n-E \frac{A_n}{\sqrt{n}}\right]\\
\ds 
+\sin \Phi_n\left[ (n+1)\frac{A_{n+1}}{\sqrt{n+1}}\cos \alpha_n+ n\frac{A_{n-1}}{\sqrt{n-1}}\cos \alpha'_n-E\frac{A_n}{\sqrt{n}}\right]\;=\;0.
\end{array}
\end{equation}
Expressions in the square brackets are the series in $1/n$. Coefficients $\cos \Phi_n$ and $\sin \Phi_n$ are irregular. Therefore, (\ref{coefficients}) can be satisfied if and only if:
\begin{eqnarray}\label{condition-of-coefficients}
 (n+1)\frac{A_{n+1}}{\sqrt{n+1}}\sin \alpha_n+n\frac{A_{n-1}}{\sqrt{n-1}}\sin \alpha'_n-E\frac{A_n}{\sqrt{n}}\;=\;0;  \nonumber \\
(n+1)\frac{A_{n+1}}{\sqrt{n+1}}\cos \alpha_n+n\frac{A_{n-1}}{\sqrt{n-1}}\cos \alpha'_n-E\frac{A_n}{\sqrt{n}}\;=\;0.
\end{eqnarray}
Each LHS of (\ref{condition-of-coefficients}) is well defined series in $x=1/n$. They must be zero, so that  each coefficient in $x=1/n$ expansion must be zero. Thus (\ref{condition-of-coefficients}) provides a set of algebraic equations for $\delta_j, \epsilon_j$.

Precise form of the asymptotic corrections is the following:
\begin{equation}\label{corrections-amplitude}
\begin{array}{l}
\ds \delta(n,E)\;=\;-\frac{1}{4\,n} + \frac{2E^2+1}{32\,n^2} - \frac{5(2E^2-1)}{128\,n^3} + \frac{20E^4-60E^2-21}{2048\,n^4} \\ [5mm]
\ds -\frac{180E^4-1380E^2+399}{8192\, n^5} + \frac{120E^6-2540E^4+2518E^2+869}{65536\, n^6} + \mathcal{O}(n^{-7})
\end{array}
\end{equation}
and
\begin{equation}\label{corrections-phase}
\begin{array}{l}
\ds\epsilon(n,E)\;=\;\frac{E}{4\,n} - \frac{E(E^2-5)}{96\,n^2} + \frac{E(E^2-9)}{96\,n^3} - \frac{E(9E^4-490E^2+341)}{15360\,n^4} \\ [5mm]
\ds +\frac{E(3E^4-190E^2+375)}{2560\,n^5} - \frac{E(15E^6-2793E^4+22169E^2-7615)}{258048\,n^6} + \mathcal{O}(n^{-7})\;.
\end{array}
\end{equation}
The correction terms $\delta_j$ and $\epsilon_j$ can by produced from the recursion by a bootstrap up to any order of $1/n$.

\section{Orthogonality}

There is a remarkable way to derive the inner product for two states in our model. Consider a truncated state, 
\begin{equation}
|\psi^{(N)}_{E}\rangle \;=\; \sum_{n=0}^{N} |n\rangle \psi_n(E)\;,
\end{equation}
where $\psi_n(E)$ are defined by (\ref{recursion}) with the initial condition $\psi_0\;=\;1$. Straightforward computation gives
\begin{equation}
\boldsymbol{H}\;|\psi^{(N)}_E\rangle \;=\; |\psi^{(N-1)}_E\rangle \;E \;+\;
|N\rangle\; N\psi_{N-1}(E) \;+\; |N+1\rangle \; (N+1)\psi_N(E)\;.
\end{equation}
Considering then
\begin{equation}
\langle\psi^{(N)}_{E'}|\;\boldsymbol{H}\;|\psi^{(N)}_E\rangle\;,
\end{equation}
one deduces
\begin{equation}
\langle \psi^{(N-1)}_{E'}|\psi^{(N-1)}_E\rangle \;=\; 
\frac{N}{E-E'} \; \left(\psi_N(E)\psi_{N-1}(E')-\psi_N(E')\psi_{N-1}(E)\right)\;.
\end{equation}
Assuming our asymptotic for $\psi_N$ for $N\to\infty$, one obtains
\begin{equation}\label{inner-product}
\langle\psi^{(N)}_{E'}|\psi^{(N)}_E\rangle \;=\; A(E')A(E) \frac{\ds\sin\left(\frac{E'-E}{2}\log N + \varphi(E')-\varphi(E)\right)}{E'-E}\;,\quad N\to\infty\;.
\end{equation}
The limit $N\to\infty$ is well defined here. In general,  this is the Fresnel integral limit \cite{EG},
\begin{equation}
\lim_{K\to\infty}\frac{\sin(Kx)}{x}\;=\;\pi\delta(x)\;.
\end{equation}
Therefore, at $N\to\infty$ one obtains
\begin{equation}
\langle\psi_{E'}|\psi_{E}\rangle\;=\;\pi A(E)^2 \delta(E-E')\;.
\end{equation}
In fact, this is the main result of our paper. Numerical analysis also shows that the spectrum is unbounded since
\begin{equation}
A(E)\;=\;A(-E)\;.
\end{equation}

\section{Conclusion and Discussion}

In this paper we have considered the stationary Schr\"odinger equation for the self-conjugated Hamiltonian $\boldsymbol{H}\; =\; \frac{1}{\ii} (\eop+\fop)$, where $\eop$ and $\fop$ are creatinig and annihilation operators for the algebra $sl_2$ considered for the infinite-dimensional representation with lowest weight equals $1$, equivalent to the usual Fock Space.

The eigenvector equation for operator $\boldsymbol{H}$ is the the second order recursion equation. In this paper we have given detailed analysis for a solution of the recursion. General expression of $\psi_n(E)$ involves four functions: $ A(E)$ , $\psi(E)$, $\delta_n(E)$, $\varepsilon_n(E)$, see equation (\ref{sequences}). We give the rigorous way to define $\delta(n,E)$ and $\epsilon(n,E)$ analytically in the forms of series expansion with respect to $1/n$ and $E$, however the functions $A(E)$ and $\psi(E)$ are defined only numerically for real $E$. 

The further development of the problem implies two ways: the first way is the further analysis of equation (\ref{recursion}) in order to find analytical expressions for the asymptotic analytical functions $A(E)$ and $\varphi(E)$. The second way could be $q\neq 1$ generalisation  of the problem. A preliminary analysis shows that $q\neq 1$ case leads to several unexpected mathematical phenomena.

\section{Acknowledgement}

I would like to thank my supervisor, Prof. Sergey Sergeev, for the patient guidance, encouragement and advices he has provided throughout my time as his student. I have been extremely lucky to have a supervisor who cared so much about my work, and who responded to my questions and queries so
promptly. Also,I would like to express my special thanks of gratitude to my parents for their support and encouragement. Otherwise, my thanks and appreciation to all my brothers , sisters and my wife who always try to make me happy. Really,I am very grateful to the University of Tabuk which gave me full scholarship.


\end{document}